\documentstyle[12pt]{article}
\newtheorem{theorem}{Theorem}
\newtheorem{proposition}[theorem]{Proposition}
\newtheorem{lemma}[theorem]{Lemma}
\newtheorem{corollary}[theorem]{Corollary}
\newtheorem{definition}[theorem]{Definition}
\newtheorem{remark}[theorem]{Remark}

\def\R{{\bf R}} 
 
\def\N{{\bf N}}
  
\def\C{{\bf C}}

\def\be{\begin{equation}}
\def\ee{\end{equation}}

\def\ds{\displaystyle}

\def\re{{\rm Re}}
\def\im{{\rm Im}}

\date{}
\begin{document}
\baselineskip=18pt
\title{Spectral theory and distributional Borel summability for the quantum H\'enon-Heiles
model}

\author{
Emanuela Caliceti
 \\ \small{Dipartimento di Matematica, Universit\`{a} di Bologna
40127 Bologna, Italy}
\\ {\footnotesize and} \small{INFN, sezione Bologna}\\
{\footnotesize (caliceti@dm.unibo.it)}}
\maketitle 
\vskip 12pt\noindent 
\begin{abstract}
 { \noindent
The Borel summability in the distributional sense is established of the divergent
perturbation theory for the ground state resonance of the quantum H\'enon-Heiles
model.}
\end{abstract}
\vskip 12pt\noindent 
%%%%%%%%%%%%%%%%%%%%%%%%%%%%%%%%%%%%%%%%%%%%%%%%%%%%%%%%%%%%%%%%%%%%
%%%%%%%%%%%%%%%%%%%%%%%%%%%%%%%%%%%%%%%%%%%%%%%%%%%%%%%%%%%%%%%%%%%% 
\section{Introduction and statement of the results}
\setcounter{equation}{0}%
\setcounter{theorem}{0}% 
%%%%%%%%%%%%%%%%%%%%%%%%%%%%%%%%%%%%
%%%%%%%%%%%%%%%%%%%%%%%%%%%%%%%%%%%%
A standard model for transition to chaos is given by the H\'enon-Heiles 
Hamiltonian (see e.g. \cite{HH}, \cite{CF}, \cite{LL}) defined by
\be \label{eq1.1}
H(\beta)=p_1 ^2 +p_2 ^2 +q_1 ^2 +q_2 ^2 +\beta \left(q_1 ^2 q_2 -\frac{1}{3}q_2
^3 \right),\quad \beta \in \R .
\ee
The quantum counterpart of (\ref{eq1.1}) (see e.g. \cite{FMP},
\cite{BDMS},
\cite{BR}, \cite{CS}, \cite{BBLM}, \cite{FvML}) is represented by the
Schr\"odinger operator in $L^2 (\R^2)$ formally given by 
\be \label{eq1.2}
H(\beta)=-\Delta +x^2 +\beta \left(x_1 ^2 x_2 -\frac{1}{3}x_2 ^3 \right):=H(0)+\beta V,
\ee
where $x=(x_1,x_2)\in \R^2,\, x^2 =|x|^2 =x_1 ^2 +x_2 ^2 ,\, \Delta$ is
the $2-$dimensional Laplace operator, $V$ is the multiplication operator
by the function $V(x)=x_1 ^2 x_2 -\frac{1}{3}x_2 ^3$ and
$H(0)$ is the operator corresponding to the harmonic oscillator. Its
spectral properties have been extensively investigated numerically (see
e.g. \cite{D}). However the mathematical analysis of the problem is
somewhat tricky. The purpose of this paper is to solve one of the
mathematical problems involved, namely the meaning of perturbation
theory, as announced in \cite{Pr}, in order to obtain results analogous
to those obtained in \cite{CGM} and \cite{C-2000} for the one dimensional
odd anharmonic oscillator. Indeed, as in the one dimensional case, the
minimal operator generated by (\ref{eq1.2}), with
$C_0 ^{\infty}(\R^2)$ as domain, is not essentially selfadjoint;
in fact, it has infinitely many selfadjoint extensions, none of which
with discrete spectrum. Therefore the numerically observed
eigenvalues have to be interpreted as (real part of) resonances
\cite{N}, which are complex objects: the real part is the location, and the imaginary part
the width. Another singularity in this problem is due to the fact that the
Rayleigh-Schr\"odinger perturbation expansion (from now on denoted RSPE) near the lowest
unperturbed eigenvalue
$E_0 =2$ not only diverges but has real coefficients with constant signs. As is
well known, this prevents the series from being Borel summable in the ordinary sense.
The aim of this paper is then to prove that, as in the Stark effect (see
\cite{Stark}), the perturbation series near
$E_0$ is Borel summable in the distributional sense to the  real part of
the resonance. On the other hand, if $\beta$ is taken to be purely
imaginary, then $H(\beta)$ shows a less singular bahavior (see \cite{N} and
\cite{CG1}). In fact $H(\beta)$ is closable on $C_0 ^{\infty}(\R^2)$ and, if
$D(H(\beta))$ denotes the domain of its closure, then $H(\beta)$ is
$\mathcal{PT}$-symmetric, i.e.
$$
\overline{(H(\beta)u)(-x)}=H(\beta)\overline{u(-x)},\quad \forall u\in
D(H(\beta)).
$$
Furthermore the RSPE near any eigenvalue 
has real coefficients with alternating signs and is Borel summable to
the corresponding (real) eigenvalue of $H(\beta)$ (see \cite{N}). Thus,
the natural way of dealing with $H(\beta)$ is to start with $\beta$
complex and then look for a continuation to $\beta$ real. More
precisely we will prove the following. 
%%%%%%%%%%%%%%%%%%%%%
\begin{theorem} \label{T1.1}
\begin{itemize}
\item[(a)] $H(\beta)$ represents an analytic family of type $A$ of closed
operators with compact resolvents, with domain
$$
D(H(\beta))=D(H(0))\cap D(V),\quad \mbox{for}\;\; 0<\arg \beta<\pi . 
$$  
\item[(b)] Let $E_0 =2$ denote the ground state energy level of $H(0)$.
Then for any $\delta >0$ there is $B(\delta)>0$ such that for
$|\beta|<B(\delta),\, 0<\arg \beta <\pi ,\, H(\beta)$ has exactly one eigenvalue
$E(\beta)$ near $E_0$, which admits an analytic continuation across the
real axis to the sector
\be \label{eq1.3}
S_{\delta}:=\left\{\beta \in \C :\, 0<|\beta|<B(\delta),\;
-\frac{\pi}{4}+\delta <\arg \beta<\frac{5}{4}\pi -\delta \right\}.
\ee
Moreover, $\lim _{\stackrel{\beta \to 0}{\beta \in
S_\delta}}E(\beta)=E_0$
\end{itemize}
\end{theorem}
%%%%%%%%%%%%%%%%%%%%%
In \cite{N} weaker results concerning the spectral properties of
$H(\beta)$ are obtained, which are not sufficient to
establish our main result, stated in Theorem \ref{T1.2} below,
i.e. the distributional Borel summability (from now on denoted DBS) of the RSPE around
$E_0$. More precisely in order to specify the domain of $H(\beta)$ and guarantee the
compacteness of its resolvents (as stated in Theorem \ref{T1.1}-(a)), we need a
quadratic estimate on $H(\beta)$, proved in Appendix B.
We can now state the main result of this paper.
%%%%%%%%%%%%%%%%%%%%%
\begin{theorem} \label{T1.2}
Let $\beta \in \R$. Then 
\begin{itemize}
\item[(a)] the RSPE near $E_0$ is Borel-Leroy summable of order
$\frac{1}{2}$ in the ordinary sense to $E(\beta)$ for $0<\arg \beta<\pi$
and in the distributional sense to $\Re \, E(\beta)$ for $\beta \in \R, |\beta|$
suitably small; 
\item[(b)] $\Re \, E(\beta)=\Re  \, E(-\beta),\, \Im \, E(\beta)=-\Im \,
E(-\beta),\, \mbox{for}\,\, \beta \in \R$. 
\end{itemize}
\end{theorem}
%%%%%%%%%%%%%%%%%%%%%
The notion of distributional Borel-Leroy summability and the
corresponding criterion are recalled in Appendix A for the
convenience of the reader.
%%%%%%%%%%%%%%%%%%%%%
\begin{remark} \label{R1.3}
{\rm In order to extend the result stated in Theorem \ref{T1.2} to any
unperturbed eigenvalue $E_l =2(l+1),\, l=1,2,\ldots$ of $H(0)$, one needs
to extend the definition (and corresponding criterion) of distributional
Borel summability to the degenerate case, as Hunziker and Pillet did in \cite{HP}
for the notion of ordinary Borel summability. We plan to do this in a forthcoming
paper.}
\end{remark}
%%%%%%%%%%%%%%%%%%%%%
The proof of Theorem \ref{T1.1} is obtained in Section 2,
following \cite{C-2000} where analogous results were obtained for the
one dimensional add anharmonic oscillator. We will include most
of the details in order to make the paper self-contained. The
proof of Theorem \ref{T1.2} requires the verification of the
analogue of the Nevanlinna criterion for the DBS, as stated
in Appendix A and proved in \cite{DBS}. In particular we need to
extend the analyticity of $E(\beta)$ to a suitable Nevanlinna
disc. This result is achieved in Section 3 by using a
refinement of the Hunziker-Vock stability technique introduced in
\cite{Stark}, \cite{DW} and \cite{C-2000} where analogous results are
obtained for the resonances of the Stark effect, of
double-well oscillators and of the $1-$dimensional odd
anharmonic oscillator, respectively. However such technique
cannot be directly applied to the present $2-$dimensional
problem. Indeed, the method is based on a control of the
numerical range of $H(\beta)$, whose distance from any
complex number $z\notin \sigma (H(0))$ must be bounded
from below by a positive constant as $|x|\to \infty$ and
$\beta \to 0$. It is possible, however, to overcome this
difficulty by passing to polar coordinates, where the
angular coordinate can be easily dealt with and the
problem essentially reduces to a $1-$dimensional one.

%%%%%%%%%%%%%%%%%%%%%%%%%%%%%%%%%%%%%
%%%%%%%%%%%%%%%%%%%%%%%%%%%%%%%%%%%%%
\section{The operator $H(\beta)$ for $\Im \, \beta \ne 0$ and the analytic
continuation of the eigenvalues}
\setcounter{equation}{0}%
\setcounter{theorem}{0}% 
%%%%%%%%%%%%%%%%%%%%%%%%%%%%%%%%%%%%%
%%%%%%%%%%%%%%%%%%%%%%%%%%%%%%%%%%%%%

In order to prove Theorem \ref{T1.1} we need some preliminary results.
%%%%%%%%%%%%%%%%%%
\begin{lemma} \label{L2.1}
Let $\beta =|\beta|e^{i\alpha}$ with $\alpha \in ]0,\pi[$ and $\Omega$ be
a compact subset of
$$
\{\gamma \in \C \setminus \{0\}:|\gamma|^2 \geq 4|\beta|\sin \alpha ,\;
-\pi +\alpha <\arg \beta <\alpha \}.
$$
Then there exist $a,b>0$ such that
\be \label{eq2.1}
\Vert \Delta u\Vert ^2 +|\gamma|^2 \Vert x^2 u\Vert ^2 +|\beta|^2 \Vert
Vu\Vert ^2 \leq a\Vert (-\Delta +\gamma x^2+\beta V)u\Vert ^2 +b\Vert
u\Vert ^2
\ee
$\forall u\in C_0 ^{\infty}(\R^2),\, \gamma \in \Omega,\, 0<|\beta|\leq 1$,
$a$ and $b$ independent of $\gamma$ in $\Omega$ and $\alpha$ in a closed
interval contained in $]0,\pi[$.
\end{lemma}
%%%%%%%%%%%%%%%%%%
{\it Proof.} See Appendix B.
%%%%%%%%%%%%%%%%%%
\begin{corollary} \label{C2.2}
Let $\gamma ,\beta \in \C$ satisfy the conditions of Lemma \ref{L2.1}.
Then the operator $T(\gamma,\beta)$ defined in $L^2 (\R^2)$ by 
$$
T(\gamma,\beta)u=-\Delta u+\gamma x^2 u+\beta Vu,\quad \forall u\in
D(T(\gamma,\beta))
$$
on the domain $D(T(\gamma,\beta))=D(H(0))\cap D(V)$ is closed and has $C_0
^{\infty}(\R^2)$ as a core.
\end{corollary}
%%%%%%%%%%%%%%%%%%
%%%%%%%%%%%%%%%%%%
\begin{lemma} \label{L2.3}
Let $\gamma,\beta \in \C$ satisfy the conditions of Lemma \ref{L2.1},
i.e. $\alpha =\arg \beta \in ]0,\pi[,\, -\pi +\alpha <\arg \gamma <\alpha
,\, |\gamma|^2 >4|\beta|\sin \alpha$.\\
Then there exists $\xi >0$ such that
\be \label{eq2.2}
\xi \Re \, [e^{-i(\alpha -\pi /2)}<u,T(\gamma,\beta)u>]\geq <u,-\Delta u>\quad \forall
u\in D(T(\gamma,\beta)).
\ee
\end{lemma}
%%%%%%%%%%%%%%%%%%
{\it Proof.} It is enough to prove (\ref{eq2.2}) for $u\in C_0
^{\infty}(\R^2)$. We have
\begin{eqnarray*}  
\lefteqn{ \Re \, [e^{-i(\alpha -\pi /2)}<u,(-\Delta +\gamma x^2 +\beta V)u>] }&&\\
&&=\cos \left(\alpha -\frac{\pi}{2}\right)<u,-\Delta u>+|\gamma|\cos
\left(\frac{\pi}{2}-\alpha +\arg \gamma\right)<u,x^2 u>\\
&&=\sin \alpha <u,-\Delta u>+|\gamma|\sin (\alpha -\arg \gamma)<u,x^2 u>\geq \sin
\alpha <u,-\Delta u>
\end{eqnarray*} 
since $\sin (\alpha -\arg \gamma)>0$ because $0<\alpha -\arg \gamma <\pi$
by assumption. Now, since $0<\alpha <\pi$, the lemma is proved with
$\xi =(\sin \alpha)^{-1}$.
%%%%%%%%%%%%%%%%%%
\begin{corollary} \label{C2.4}
The numerical range of $T(\gamma,\beta)$ is contained in the half-plane
$\{z\in \C:\, -\pi +\alpha \leq \arg z\leq \alpha \}$.
\end{corollary}
%%%%%%%%%%%%%%%%%%
Set $H(\beta):=T(1,\beta)$, for $\alpha =\arg \beta \in ]0,\pi[$. By
the above results we have the following theorem, which corresponds to
Theorem \ref{T1.1}-(a). 
%%%%%%%%%%%%%%%%%%
\begin{theorem} \label{T2.5}
$H(\beta)$ represents an analytic family  of type $A$ of operators with compact
resolvents, with $D(H(\beta))=D(H(0))\cap D(V)$, for $0<\arg \beta <\pi$. 
\end{theorem}
%%%%%%%%%%%%%%%%%%
We will prove that the (discrete) spectrum of $H(\beta)$ is non-empty (see also
\cite{N} ). In order to prove Theorem \ref{T1.2}, i.e. the DBS of the RSPE near $E_0$
we need to prove that it is stable with respect to the family $H(\beta)$, as
$\beta
\to 0, \Im \beta >0$ and that the corresponding eigenvalue
$E(\beta)$ of $H(\beta)$ can be analytically continued to a wider
sector than $0<\arg \beta <\pi$. To this end we start by making use of standard
dilation analyticity techniques, i.e. we introduce the operator
\be \label{eq2.3}
H(\beta,\theta):=-e^{-2\theta}\Delta +e^{2\theta}x^2 +\beta
e^{3\theta}V(x):=e^{-2\theta}K(\beta,\theta)
\ee    
which, for $\theta \in \R$, is unitarily equivalent to $H(\beta),\Im \beta >0$, via
the dilation operator
$$
(U(\theta)u)(x)=e^{\theta}u(e^{\theta}x),\qquad \forall u\in L^2 (\R^2)
$$
First of all notice that $K(\beta,\theta)=-\Delta +e^{4\theta}x^2 +\beta
e^{5\theta}V(x)$ corresponds to $T(\gamma,\beta')$ with $\gamma =e^{4\theta},\beta
'=\beta e^{5\theta}$. From now on we will assume $|\Re \theta|<1$ and
$|\beta|<\frac{e^{-3}}{4}$ so that the conditions $|\beta '|<1$ and $|\gamma|^2
>4|\beta '|\sin (\arg \beta ')$ required in Lemma \ref{L2.1} are
automatically satisfied.\\
Next observe that the further conditions on $\beta '$ and $\gamma$ so far required:
\be \label{eq2.4}
\left\{ \begin{array}{ll}
-\pi +\arg \beta '<\arg \gamma <\arg \beta ' \\
0<\arg \beta '<\pi
\end{array} \right.
\ee
are equivalent to the following conditions on $\beta$ and $\theta$:
\be \label{eq2.5}
\left\{ \begin{array}{ll}
0<\arg \beta +\Im \theta <\pi \\
0<\arg \beta +5\Im \theta<\pi
\end{array} \right.
\ee
In complete analogy with Theorem \ref{T2.5} of \cite{C-2000} we can now prove the
following
%%%%%%%%%%%%%%%%%%%%
\begin{theorem} \label{T2.6}
Let $s=\arg \beta ,t=\Im \theta$. Then $H(\beta,\theta)$ is a holomorphic family
of type $A$ of closed operators on $D(H(\beta,\theta))=D(H_0)\cap D(V)$ with
compact resolvents for $\beta$ and $\theta$ such that $s$ and $t$ vary in the
parallelogram $P$ of the $(s,t)-$plane defined by 
$$
P=\{(s,t)\in \R^2 :\, 0<t+s<\pi ,0<5t+s<\pi \}\, .
$$  
\end{theorem}
%%%%%%%%%%%%%%%%%%%%
%%%%%%%%%%%%%%%%%%%%
\begin{remark} \label{R2.7}
{\rm \begin{enumerate}
\item By Corollary \ref{C2.4}, the numerical range of $K(\beta,\theta)$ is
contained in the half-plane $-\pi +\alpha \leq \arg z\leq \alpha$ with $\alpha
=\arg \beta +5\Im \theta$; thus, $H(\beta,\theta)$ has numerical range contained
in the half-plane
\be \label{eq2.6}
\Pi =\{z\in \C :\, -\pi +\arg \beta +3\Im \theta \leq \arg z \leq \arg \beta +3
\Im \theta \}\, .
\ee
Moreover the (discrete) spectrum of $H(\beta,\theta)$ is contained in $\Pi$ and
$\forall z\notin \Pi ,\, \Vert (z-H(\beta,\theta))^{-1}\Vert \leq \mbox{dist}\,
(z,\Pi)^{-1}$. Finally, the analyticity of $H(\beta,\theta)$ in the region
defined by $P$ allows $\beta$ to be extended to the sector $S=\{\beta :\,
0<|\beta|<\beta _0 ,-\frac{\pi}{4}<\arg \beta <\frac{5}{4}\pi \}$
\item If we start from the operator $H(\beta)$ with $\Im \beta <0$, analogous
results can be obtained for the operator family $H(\beta,\theta)$ for $\beta$ and
$\theta$ satisfying the conditions:
\be \label{eq2.7}
\left\{ \begin{array}{ll}
-\pi <\Im \theta +\arg \beta <0 \\
-\pi <5\Im \theta +\arg \beta <0
\end{array} \right.
\ee 
Furthermore $H(\beta,\theta)^* =H(\bar \beta,\bar \theta)$.
\end{enumerate} }
\end{remark}
%%%%%%%%%%%%%%%%%%%%
By standard dilation analyticity arguments the eigenvalues $E_l =2(l+1),\,
l=0,1,\ldots$, of $H(0,\theta)=-e^{-2\theta}\Delta +e^{2\theta}x^2
,D(H(0,\theta))=D(H(0))$, are independet of $\theta$ for $-\frac{\pi}{4}<\Im
\theta <\frac{\pi}{4}$. By an argument similar to that used to prove Theorem
2.7 in \cite{C-2000} (see also \cite{N}) we can prove the stability in the sense of
Kato of each eigenvalue $E_l$ with respect to the family
$\{H(\beta,\theta):\,|\beta|>0\}$, $\beta$ and $\theta$ in the region defined
by $P$. More precisely we can state the following 
%%%%%%%%%%%%%%%%%%
\begin{theorem} \label{T2.8}
Let $\beta$ and $\theta$ satisfy conditions (\ref{eq2.5}). We have
\begin{itemize}
\item[(a)] if $\lambda \notin \sigma (H(0,\theta))$, then $\lambda \in \cal D$,
where
$$
{\cal D}=\{z\in \C:\,z\notin \sigma (H(\beta,\theta))\,\mbox{and}\,
(z-H(\beta,\theta))^{-1}\,\mbox{is uniformly bounded as}\,|\beta|\to 0\}
$$
\item[(b)] if $\lambda \in \sigma (H(0,\theta))=\{2(l+1):\, l=0,1,\ldots \}$,
then $\lambda$ is stable with respect to the family $H(\beta,\theta)$, i.e. if
$r>0$ is sufficiently small, so that the only eigenvalue of $H(0,\theta)$
enclosed in $\Gamma _r =\{z\in \C:\, |z-\lambda|=r\}$ is $\lambda$, then there
is $B>0$ such that for $|\beta|<B,\dim P(\beta,\theta)=\dim P(0,\theta)$, where
$$
P(\beta,\theta)=(2\pi i)^{-1}\oint _{\Gamma _r}(z-H(\beta,\theta))^{-1}dz
$$  
is the spectral projection of $H(\beta,\theta)$ corresponding to the part of the
spectrum enclosed in $\Gamma _r \subset \C \setminus \sigma
(H(0,\theta))$.
\end{itemize} 
\end{theorem}
%%%%%%%%%%%%%%%%%%
%%%%%%%%%%%%%
\begin{remark} \label{R2.9}
{\rm It can be immediately checked that all the results so far obtained, in
particular the analyticity of the family $H(\beta,\theta)$ and the stability of
the eigenvalues of the harmonic oscillator $H(0,\theta)$ with respect to
$H(\beta,\theta)$ as $|\beta|\to 0$, hold unifomly in $\beta$ and $\theta$ such that
$(\arg \beta ,\theta)$ varies in any compact subset of
$P$}.
\end{remark}
%%%%%%%%%%%%%
Now we specialize the result obtained in Theorem \ref{T2.8} to the ground
state energy level $E_0 =2$ of $H(0)$. More precisely, for any $\delta >0$
there exists $B(\delta)>0$ such that for
$|\beta|<B(\delta),-\frac{\pi}{4}+\delta<\arg \beta <\frac{5}{4}\pi -\delta
,H(\beta,\theta)$ has one and only one eigenvalue $E(\beta)$, independent of
$\theta$ if $(\arg \beta,\Im \theta)\in P$, which converges to $E_0$ as
$|\beta|\to 0$.\\
By Theorem \ref{T2.6}, $E(\beta)$ is analytic in the sector
\be \label{eq2.8}
S_{\delta}=\left\{\beta \in \C:\, 0<|\beta|<B(\delta),\,-\frac{\pi}{4}+\delta<\arg
\beta <\frac{5}{4}\pi -\delta \right\}
\ee
and is an eigenvalue of $H(\beta)$ for $0<\Im \beta<\pi$. For future reference we
state this result in the following
%%%%%%%%%%%%%%%%%%%
\begin{theorem} \label{T2.10}
For any $\delta >0$, there is $B(\delta)>0$ such that for
$|\beta|<B(\delta),0<\arg \beta <\pi$, $H(\beta)$ has exactly one eigenvalue
$E(\beta)$ near $E_0$, which admits an analytic continuation across the real axis
to the sector $S_\delta$. Moreover $\lim _{\stackrel{\beta \to 0}{\beta \in
S_\delta}}E(\beta)=E_0$. 
\end{theorem}
%%%%%%%%%%%%%%%%%%%

%%%%%%%%%%%%%%%%%%%%%%%%%%%%%%%%%%%%%
%%%%%%%%%%%%%%%%%%%%%%%%%%%%%%%%%%%%%
\section{Analyticity of $E(\beta)$ in a Nevanlinna disc and DBS}
\setcounter{equation}{0}%
\setcounter{theorem}{0}% 
%%%%%%%%%%%%%%%%%%%%%%%%%%%%%%%%%%%%%
%%%%%%%%%%%%%%%%%%%%%%%%%%%%%%%%%%%%%

The basic analyticity result needed to establish the DBS  of the RSPE near $E_0$ for
$\beta >0$ is obtained in the following theorem.
%%%%%%%%%%%%%%%%%%%
\begin{theorem} \label{T3.1}
There exists $R>0$ such that the eigenvalue $E(\beta)$ of
$H(\beta)=-\Delta +x^2 +\beta (x_1 ^2 x_2 -\frac{1}{3}x_2 ^3)$ near $E_0$ for
$|\beta|$ small is analytic in the Nevanlinna disc $C_R =\{\beta :\, \Re \beta
^{-2}>R^{-1}\}$ in the
$\beta ^2 -$plane.
\end{theorem}
%%%%%%%%%%%%%%%%%%%
%%%%%%%%%%%%%%%%
\begin{remark} \label{R3.2}
\begin{itemize}
\item[{\rm (I)}] {\rm The sector $S(\delta)$ given by (\ref{eq2.8}) can be
rewritten in terms of the variable $\beta ^2$ as:}
\be \label{eq3.1}
S(\delta)=\left\{\beta :\,0<|\beta|<B(\delta),\,-\frac{\pi}{2}+2\delta <\arg
\beta ^2 <\frac{\pi}{2}+2\pi -2\delta \right\}\, .
\ee
\item[{\rm (II)}] {\rm The function $E(\beta)$, analytic in any sector $S(\delta)$
and for which we want to prove analyticity in $C_R$, represents an eigenvalue of
$H(\beta,\theta)$ if $(\arg \beta,\Im \theta)\in P$. In particular, for
$-\frac{\pi}{4}<\arg \beta <0$ we can choose the path inside $P$ given by the
straight line of equation}
$$
\Im \theta =-\frac{1}{3}\arg \beta +\frac{\pi}{6}\, .
$$ 
{\rm Then, if we set}
\be \label{eq3.2}
\arg \beta =-\frac{\pi}{4}+\frac{\varepsilon}{2},\;\; \mbox{{\rm i.e.}}\;\arg \beta
^2 =-\frac{\pi}{2}+\varepsilon,\;\; \varepsilon \to 0^+
\ee
{\rm we obtain $\Im \theta =\frac{\pi}{4}-\frac{\varepsilon}{6}$, and the operator
$H(\beta,\theta)$ takes the form}
\be \label{eq3.3}
A(\rho)=-\theta _0 ^{-2}e^{-i(\frac{\pi}{2}-\frac{\varepsilon}{3})}\Delta +\theta
_0 ^2 e^{i(\frac{\pi}{2}-\frac{\varepsilon}{3})}x^2 +i\rho \theta _0 ^3 V(x)
\ee
{\rm with $\rho =|\beta|$ and $\theta _0 =e^{\Re \theta}$}.
\item[{\rm (III)}] {\rm For $\beta =\rho e^{i\arg \beta}$ and $\arg \beta
=-\frac{\pi}{4}+\frac{\varepsilon}{2}$, the boundary of $C_R$ has the equation}
\be \label{eq3.4}
\sin \varepsilon =\frac{\rho ^2}{R}\, .
\ee 
\item[{\rm (IV)}] {\rm Since the disc $C_R$ is the union of the boundaries of discs
of smaller radius, the proof of Theorem \ref{T3.1} reduces to a stability argument for
the eigenvalue $E_0$ with respect to the family $\{A_{\rho}\}_{\rho >0}$ as $\rho \to
0^+$. In view of Remark 3.2 (IV) it is convenient to move to polar coordinates
$(r,\varphi)$ as follows:
\be \label{eq3.5}
\left\{ \begin{array}{ll}
x_1 =r\cos \varphi \\
x_2 =r\sin \varphi
\end{array}\right.
\ee
Then $A(\rho)$ is equivalent to the operator $A_1 (\rho)$ formally given by
\begin{eqnarray} \label{eq3.6}
A_1(\rho) & = & \theta _0
^{-2}e^{-i(\frac{\pi}{2}-\frac{\varepsilon}{3})}\left\{-\frac{\partial
^2}{\partial r^2}-\frac{1}{r^2}\frac{\partial ^2}{\partial \varphi
^2}-\frac{1}{4r^2}\right\}+\theta _0 ^2
e^{i(\frac{\pi}{2}-\frac{\varepsilon}{3})}r^2 {} 
                                             \nonumber\\
& & {}+i\rho \theta _0 ^3 r^3 \left(\cos ^2 \varphi \sin \varphi -\frac{1}{3}\sin ^3
\varphi \right)
\end{eqnarray}
in $L^2 (\R^+ \times {\mathbf T}^1)$ with suitable conditions at the origin.\\
More precisely $A_1 (\rho)$ is the operator generated by the quadratic form
\begin{eqnarray} \label{eq3.7}
a_{\rho}[u] & := & \int _0 ^{\infty}\int _0 ^{2\pi}\Bigg\{\theta _0
^{-2}e^{-i(\frac{\pi}{2}-\frac{\varepsilon}{3})}\left[|p_r u|^2
+\frac{1}{r^2}|p_{\varphi}u|^2 -\frac{1}{4r^2}|u|^2\right]
                                                          \nonumber\\
& & +\left[\theta _0 ^2 e^{i(\frac{\pi}{2}-\frac{\varepsilon}{3})}r^2 +i\rho \theta _0
^3 r^3 \left(\cos ^2 \varphi \sin \varphi -\frac{1}{3}\sin ^3
\varphi \right)\right]|u|^2\Bigg\}drd\theta \nonumber\\
& & 
\end{eqnarray}
defined on the maximal domain with the condition at the origin $u(r,\varphi)\simeq
r^{1/2}$ uniformly in $\varphi$. Here we adopt the notation $p_r
=-i\frac{\partial}{\partial r},p_{\varphi}=-i\frac{\partial}{\partial
\varphi}$. Now, as in \cite{Stark}, let $\cal U$ be a transformation in the set of
$L^2$ functions which are translation analytic in some suitable strip $|\Im
r|<\delta _0$, defined by:
\be \label{eq3.8}
({\cal U}\psi)(r,\theta)=\xi'(r)^{1/2}\psi (\xi (r),\theta)  
\ee 
where, setting $r_0 =\frac{a_0}{\rho}$ for a suitable $a_0 >0$, we define
$\xi (r)\in C_0 ^{\infty}(\R^+)$ so that, for $\eta _0 \in ]0,1[$ fixed,
\be \label{eq3.9}
\xi (r)=r-2i\eta _0 [1-(1+r^3)^{-1/6}],\qquad 0<r\leq r_0
\ee
\be \label{eq3.10}
\xi (r)=r,\qquad r\geq r_0 +\eta _0
\ee
Setting $f(r)=\frac{1}{\xi'(r)}$ and $\omega
=e^{-i(\frac{\pi}{2}-\frac{\varepsilon}{3})}$, the transformed operator ${\cal U}
A(\rho){\cal U}^{-1}$ is given by
\begin{eqnarray} \label{eq3.11}
H_{\rho} & = & \omega \theta _0 ^{-2}\left[p_r f^2 p_r
+\frac{1}{4}(f^2)''-\frac{1}{\xi ^2}p_{\varphi}^2 -\frac{1}{4\xi ^2}\right]{}
                                                                     \nonumber\\
& & {}+\omega ^{-1}\theta _0 ^2 \xi ^2 +i\rho \theta _0 ^3 \xi ^3 \left(\cos ^2 \varphi
\sin
\varphi -\frac{1}{3}\sin ^3 \varphi \right) \, .
\end{eqnarray}
The quadratic form which ganerates $H_\rho$ is defined by: 
\begin{eqnarray} \label{eq3.12}
h_{\rho}[u] & = & \int _0 ^{\infty}\int _0 ^{2\pi}\Bigg\{\omega \theta _0 ^{-2}\Bigg[f^2
|p_r u|^2 +\frac{1}{\xi ^2}|p_{\varphi}u|^2 \Bigg]
                                                \nonumber\\
& &\quad+\Bigg[\frac{\omega \theta _0 ^{-2}}{4}(f_2)''-\frac{\omega \theta _0
^{-2}}{4\xi ^2}+\omega ^{-1}\theta _0 ^2 \xi ^2
                                             \nonumber\\
& &\quad+i\rho \theta _0 ^3 \xi ^3 \Bigg(\cos ^2
\varphi \sin \varphi -\frac{1}{3}\sin ^3 \varphi \Bigg)\Bigg]|u|^2 \Bigg\}drd\varphi
\end{eqnarray}
on the maximal domain with the condition at the origin $u(r,\varphi)\simeq
r^{1/2}$, uniformly in $\varphi$. The limit in the strong resolvent sense as
$\rho \to 0^+$ of $H_\rho$ is defined by
\be \label{eq3.13}
H_0 =-i\theta _0 ^{-2}\left\{p_r f_0 ^2 p_r +\frac{1}{4}(f_0 ^2)''-\frac{1}{4\xi _0
^2}-\frac{1}{\xi _0 ^2}p_{\varphi}^2\right\}+i\theta _0 ^2 \xi _0 ^2
\ee
where $f_0 =(\xi'_0)^{-1}$ and $\xi _0 (r)$ is defined by (\ref{eq3.9})
$\forall r>0$. From Remark \ref{R3.2}-(IV) it follows that in order to prove
Theorem \ref{T3.1} it suffices to prove the stability of the eigenvalue $E_0$ of
$H_0$ with respect to the family $H_\rho$ as $\rho \to 0^+$. As in \cite{Stark} this
result is achieved by means of some preliminary results. } 
\end{itemize}
\end{remark}
%%%%%%%%%%%%%%%
%%%%%%%%%%%%%%%%%%
\begin{lemma} \label{L3.3}
Let $\sin \varepsilon =\frac{\rho ^2}{R}$. There exists $\rho _0 >0,n_0 \in
\mathbf N$ and positive real constants $a_1 ,a_2 ,c,c_1 ,c_2$ such that 
\begin{eqnarray} \label{eq3.14}
\Re h_{\rho}[u] & \geq & a_1 \int _0 ^{2\pi}\int _1 ^{r_0}\bigg\{\sin
\left(\frac{\varepsilon}{3}\right)[1-r^4 (1+r^3)^{-7/3}]+\eta _0 r^2
(1+r^3)^{-7/6}\bigg\}|p_r u|^2 drd\varphi
                                       \nonumber\\
& &\quad+a_2 \int _0 ^{2\pi}\int _{r_0}^{\infty}\sin
\left(\frac{\varepsilon}{3}\right)|p_r u|^2 dr d\varphi -c\Vert u\Vert ^2
\end{eqnarray}
$\forall \rho \in ]0,\rho _0],\forall u\in D(h_p)$. A similar estimate holds with
$h_\rho$ replaced by $h_0 \,(r_0 =+\infty ,\varepsilon =0)$.\\
Moreover, $\forall u \in D(h_\rho)$ such that $\mbox{\rm supp}\; u \subset (n,+\infty)$
\be \label{eq3.15}
\Re h_{\rho}[u]\geq (c_1 R^{-1}-c_2)\Vert u\Vert ^2
\ee
$\forall n\geq n_0$ and $c_1 ,c_2$ independent of $R$, $\forall \rho \in [0,\rho _0]$.
\end{lemma}
%%%%%%%%%%%%%%%%%%%%
{\it Proof.} We have
\begin{eqnarray} \label{eq3.16}
\lefteqn{ \int _0 ^{2\pi}\int _0 ^{\infty}\left\{f^2 |p_r u|^2 -\frac{1}{4\xi
^{2}}|u|^2\right\}drd\varphi }&&
                                \nonumber\\
&&=\int _0 ^{2\pi}d\varphi \int _0 ^{\infty}f^2 \left[|p_r u|^2
-\frac{1}{4r^2}|u|^2\right]dr+\frac{1}{4}\int _0 ^{2\pi}d\varphi \int _0
^{\infty}\left(\frac{f^2}{r^2}-\frac{1}{\xi ^2}\right)|u|^2 dr \, .
                                                           \nonumber\\
&& 
\end{eqnarray}
Since the function $\frac{f^2}{r^2}-\frac{1}{\xi ^2}$ is bounded we obtain 
\begin{eqnarray} \label{eq3.17}
\lefteqn{ \Re \int _0 ^{2\pi}\int _0 ^{\infty}\omega \left\{f^2 |p_r u|^2 -\frac{1}{4\xi
^2}|u|^2 dr d\varphi\right\}{} }
                         \nonumber\\
& & {}\geq \int _0 ^{2\pi}\int _0 ^{\infty}\Re (\omega f^2)\left[|p_r u|^2
-\frac{1}{4r^2}|u|^2\right]drd\varphi -(\mbox{const.})\Vert u\Vert ^2
                                                            \nonumber\\
& & {}=\int _0 ^{2\pi}\int _0 ^1 (\cdots)drd\varphi +\int _0 ^{2\pi}\int _1
^{r_0}(\cdots)drd\varphi +\int _0 ^{2\pi}\int _{r_0}^{\infty}(\cdots)drd\varphi
-(\mbox{const.})\Vert u\Vert ^2 \, . 
                              \nonumber\\
& &
\end{eqnarray}
Let us now denote $I_1 ,I_2$ and $I_3$ the first, second and third integral
respectively, in the right hand side of (\ref{eq3.17}). For $r\leq r_0$ we have 
\begin{eqnarray} \label{eq3.18}
\Re f^2 & \geq & 4^{-1}\{1-r^4 (1+r^3)^{-7/3}\} \nonumber\\
\\
\Im f^2 & \geq & 2^{-1}\eta _0 r^2 (1+r^3)^{-7/6} \, . \nonumber
\end{eqnarray}  
Hence,
\be \label{eq3.19}
\Re (\omega f^2)\geq 4^{-1}\sin \left(\frac{\varepsilon}{3}\right)[1-r^4
(1+r^3)^{-7/3}]+\frac{\eta _0}{2}\cos \left(\frac{\varepsilon}{3}\right)r^2
(1+r^3)^{-7/6} \; .
\ee
Now, by Sobolev's inequality $I_1 \geq 0$; moreover
\begin{eqnarray} \label{eq3.20-21}
I_3 & \geq & \int _0 ^{2\pi}\int _0 ^{\infty}\sin
\left(\frac{\varepsilon}{3}\right)|p_r u|^2 drd\varphi -\,(\mbox{const.})\, \Vert
u\Vert^2 \, ,\\
I_2 & \geq & a_1 \int _0 ^{2\pi}\int _1 ^{r_0}\bigg\{\sin
\left(\frac{\varepsilon}{3}\right)[1-r^4 (1+r^3)^{-7/3}]\nonumber\\
& & \quad+\eta _0 r^2 (1+r^3)^{-7/6}\bigg\}|p_r u|^2 drd\varphi \, .
\end{eqnarray}
Let us now estimate the remaining terms appearing in $\Re h_\rho [u]$. We have 
\be \label{eq3.22}
\Re \int _0 ^{2\pi}\int _0 ^{\infty}\xi ^{-2}|p_{\varphi}u|^2 drd\varphi =\int _0
^{2\pi}\int _0 ^{\infty}\frac{r^2 -4\eta ^2 (r)}{(r^2 +\eta
(r)^2)^2}|p_{\varphi}u|^2 drd\varphi
\ee 
where
\be \label{eq3.23}
\eta (r)=\left\{ \begin{array}{ll}
\eta _0 [1-(1+r^3)^{-1/6}], & \mbox{for $r<r_0$}\\
0 & \mbox{for $r>r_0 +\eta _0$} \; .
\end{array} \right.
\ee
By choosing $\eta _0$ suitably small (e.g. $\eta _0 <2^{-6/5}$) we have $(r^2
-4\eta ^2)(r^2 +\eta ^2)^2 \geq 0$ and therefore (\ref{eq3.22}) is non negative.
Next notice that $(f^2)''$ is bounded; thus, 
\be \label{eq3.24}
\Re \int _0 ^{2\pi}\int _0 ^{\infty}\frac{\omega \theta _0 ^{-2}}{4}(f^2)''|u|^2
drd\varphi \geq -\,(\mbox{const.})\,\Vert u\Vert ^2 \, .
\ee   
We can now estimate the potential term $V_\rho (r,\varphi):=\alpha ^{-1}\theta
_0 ^2 \xi ^2 +i\rho \theta _0 ^3 \xi ^3 (\cos ^2 \varphi \sin \varphi
-\frac{1}{3}\sin ^3 \varphi)$. We have
\begin{eqnarray} \label{eq3.25}
\Re (V_\rho (r,\varphi)) & = & \sin \left(\frac{\varepsilon}{3}\right)(r^2 -\eta
(r)^2)\theta _0 ^2 +4\cos \left(\frac{\varepsilon}{3}\right)r\eta (r)\theta _0 ^2
                                                                              \nonumber\\
& & +6\rho r^2 \theta _0 ^3 \eta
(r)\left(\cos ^2 \varphi \sin \varphi -\frac{1}{3}\sin ^3 \varphi \right)\, . 
\end{eqnarray}
Fixing $R>0$, for any $b>0$ there exists $k>0$ such that for $r\in (0,b)$:
\be \label{eq3.26}
\Re V_\rho (r,\varphi)\geq -k \, . 
\ee
Moreover:
\begin{itemize}
\item[(I)] For $r\geq r_0 +\eta _0$, recalling that $r_0 =a_0 /\rho$, we have 
\be \label{eq3.27}
\Re V_\rho (r,\varphi)\geq k_1 \frac{\rho ^2}{R}r_0 ^2 =\frac{k_1 a_0 ^2}{R} 
\ee
for a suitable constant $k_1 >0$.
\item[(II)] Finally, for $b<r<r_0 +\eta _0$ ($b>0$, independent of $\rho >0$) we
have $0\leq \eta (b)\leq \eta (r)\leq \eta _0$. Thus, 
\begin{eqnarray} \label{eq3.28}
\Re V_\rho (r,\varphi) & \geq & -k_2 +4\cos \left(\frac{\varepsilon}{3}\right)r\theta _0
^2
\eta (r)-8r^2\rho \eta_0 \theta _0 ^3
                              \nonumber\\ 
& \geq & -k_2 +2r\eta (b)\theta _0 ^2 -8\eta _0 r^2 \rho \theta _0 ^3
\end{eqnarray}
for some $k_2 >0$, if $\rho >0$ is sufficiently small so that $\cos
(\frac{\varepsilon}{3})>\frac{1}{2}$. By suitably choosing $a_0 >0$ (e.g. $a_0
=\eta (b)/8\eta _0 \theta _0 ,b=1$) the term $2r\eta (b)-8\rho \eta _0 r^2$
attains its maximum at $r=r_0 =\frac{a_0}{\rho}$ and its minimum at $r=b$ in
$(b,r_0)$. Thus, for $r\in (b,r_0)$ we have
\be \label{eq3.29}
\Re V_\rho (r,\varphi)\geq -k_2 +2b^2 \eta (b)\theta _0 ^2 -8\rho \eta _0 b^2
\theta _0 ^3 \geq -k_2 ,
\ee
and this concludes the proof of (\ref{eq3.14}).
\end{itemize}
As for (\ref{eq3.15}) notice that the kinetic part is $\geq -\,(\mbox{const.})\,
\Vert u\Vert ^2$, and for the potential part we have:
\begin{itemize}
\item[(I')] For $r\geq r_0 +\eta _0$ we can repeat the argument used in (I).
\item[(II')] For $n_0 \leq n\leq r\leq r_0$ we proceed as in (II) with $n_0 \geq b
\, (b=1),\eta (r)\geq \eta (b)$; thus, 
\be \label{eq3.30}
\Re V_\rho (r,\varphi)\geq -k_2 +2r\eta (b)\theta _0 ^2 -8\eta _0 r^2 \rho \theta
_0 ^3\, .
\ee 
Again $2r\eta (b)\theta _0 ^2 -8\eta _0 \rho r^2 \theta _0 ^3$ attains its maximum
in $(n,r_0)$ at $r_0$ and its minimum at $r=n$. Hence
\begin{eqnarray} \label{eq3.31}
\Re V_\rho (r,\varphi) & \geq & -k_2 +2n\eta (b)\theta _0 ^2 -8n^2 \rho \eta _0 \theta
_0 ^3 \nonumber\\
& \geq & k_3 n-k_4 \geq \frac{k_5}{R}-k_6 .
\end{eqnarray}
\end{itemize}
This concludes the proof of the Lemma.
%%%%%%%%%%%%%%%%%%%%%%%%%
%%%%%%%%%%%%%%%%%%%%%%%%%
\begin{corollary} \label{C3.4}
Let $\chi _n (r)=\chi (r/n),\,\chi \in C^{\infty}(\R^+),\,\chi (r)=1$ for $r\leq
1,\chi (r)=0$ if $r\geq \frac{3}{2}$. Then there exists $c_3 >0$ such that 
\be \label{eq3.32}
\Vert [H_\rho ,\chi _n]u\Vert \leq c_3 n^{-1/4}(\Vert H_\rho u\Vert +\Vert u\Vert)\quad
\forall u\in D(H_\rho),\,0\leq \rho <\rho _0 .
\ee
\end{corollary}
%%%%%%%%%%%%%%%%%%%%%%%%
%%%%%%%%%%%%%%%%%%%%%%%%
{\it Proof.} It is enough to prove (\ref{eq3.32}) for $\rho >0$, since for $\rho
=0$ the argument is simpler. For simplicity we set $\Vert u\Vert =1$. Let $\gamma
_{2n}(r)$ be the characteristic function of the interval $[1,2n]$ in the $r$
variable. Then we have:
\be \label{eq3.33}
[H_{\rho},\chi _n]=-\theta _0 ^{-2}\gamma _{2n}\omega \left\{2in^{-1}f^2 \chi '
\left(\frac{r}{n}\right)+n^{-2}\chi '' \left(\frac{r}{n}\right)+2ff'\chi
'\left(\frac{r}{n}\right)\right\}\, . 
\ee
Notice that the term $-\frac{\omega \theta _0 ^{-2}}{\xi ^2}p_{\varphi}^2$ in
$H_\rho$ gives no contribution to the commutator $[H_\rho,\chi _n]$ since $\chi
(r)$ does not depend on $\varphi$. Now:
\begin{eqnarray} \label{eq3.34}
\Vert [H_\rho,\chi _n]u\Vert & \leq & c_4 n^{-1}\left\{\left(\int _0 ^{2\pi}\int _1
^{2n}d\varphi dr|p_r u|^2 \right)^{1/2}+1\right\}
                                               \nonumber\\
& \leq & c_5 n^{-1}\left\{\frac{(1+8n^3)^{7/12}}{2n}\left[\int _0 ^{2\pi}d\varphi \int
_1 ^{2n}dr|p_r u|^2 \eta _0 r^2 (1+r^3)^{-7/6}\right]^{1/2}+1\right\}
                                                                \nonumber\\
& \leq & c_6 n^{-1/4}\left\{\left[\int _0 ^{2\pi}d\varphi \int
_1 ^{r_0}\eta _0 r^2 (1+r^3)^{-7/6}|p_r u|^2 dr\right]^{1/2}+1\right\}
                                                                    \nonumber\\
& \leq & c_7 n^{-1/4}\{[\Re <H_{\rho}u,u>+c_8]^{1/2}+1\}
\end{eqnarray}
where the last inequality follows from Lemma \ref{L3.3}. Now, taking $c_8 =c_9 +1$
with $\Re <H_{\rho}u,u>+c_9 \geq 0$ we obtain:
\be \label{eq3.35}
\Vert [H_\rho ,\chi _n]u\Vert \leq c_{10}n^{-1/4}\{\Re <H_{\rho}u,u>+c_{11}\}
\ee
where $c_{11}=c_8 +1$. This concludes the proof.
%%%%%%%%%%%%%%%%%%%%%%
%%%%%%%%%%%%%%%%%%%%%%
\begin{remark} \label{R3.5}
{\rm A similar argument can be used to obtain the analogous estimate for the
adjoint operator $H_\rho ^*$:
\be \label{eq3.36}
\Vert [H_\rho ^* ,\chi _n]u\Vert \leq c_3 n^{-1/4}(\Vert H_\rho ^* u\Vert +\Vert
u\Vert)\quad \forall u\in D(H_\rho ^*),\quad 0\leq \rho <\rho _0 \, .
\ee }
\end{remark}
%%%%%%%%%%%%%%%%%%%%%%
%%%%%%%%%%%%%%%%%%%%%%
\begin{proposition} \label{P3.6}
Let $M_n =1-\chi _n$, where $\chi _n$ is defined as in Corollary \ref{C3.4}.\\
If
$$
d_n (\lambda ,\rho)=\inf \{\Vert (\lambda -H_\rho)M_n u\Vert :\, \Vert M_n
u\Vert =1,\, u\in D(H_\rho)\} 
$$
then $\forall \lambda \in \C,\, \exists R,n_0 ,\rho _0 ,\delta >0$ such that 
\be \label{eq3.37}
d_n (\lambda ,\rho)\geq \delta >0,\quad \forall n\geq n_0 ,\quad \forall \rho \leq
\rho _0 \, .
\ee
\end{proposition}
%%%%%%%%%%%%%%%%%%%%%%
%%%%%%%%%%%%%%%%%%%%%%
{\it Proof.} First of all notice that
\be \label{eq3.38}
d_n (\lambda ,\rho)\geq \mbox{dist}(\lambda ,E_n (\rho))
\ee
where
$$
E_n (\rho)=\{<M_n u,H_\rho M_n u>:\, u\in D(H_\rho),\,\Vert M_n u\Vert =1\}
$$
Moreover, by (\ref{eq3.15}) we have:
\be \label{eq3.39}
\Re <M_n \rho ,H_\rho M_n u>\geq c_1 R^{-1}-c_2 \, .
\ee
Now, (\ref{eq3.37}) follows from (\ref{eq3.39}): since $c_1$ and $c_2$ are
independent of $R$ we can take $R>0$ suitably small so that (\ref{eq3.37}) is
satisfied with $\delta =\frac{c_1}{R}-c_2 -|\lambda|$.
%%%%%%%%%%%%%%%%%%%%%%
%%%%%%%%%%%%%%%%%%%%%%
\begin{lemma} \label{L3.7}
Let $\rho _m >0$ and $u_m \in D(H_{\rho _m})$ be two sequences such that $\rho _m
\to 0^+ ,\Vert H_{\rho _m}u_m\Vert$ is bounded and $\Vert u_m\Vert =1,\, u_m
\stackrel{w}{\to}0$. Then the sequences $\rho _{m(n)},M_m u_{m(n)}$ satisfy the same
properties for suitable $m=m(n)$, if $R>0$ is chosen sufficiently small.
\end{lemma}
%%%%%%%%%%%%%%%%%%%%%%
%%%%%%%%%%%%%%%%%%%%%%
{\it Proof.} By Corollary \ref{C3.4} the boundedness of $\Vert H_{\rho
_m}u_m\Vert$ implies that of $\Vert H_{\rho _{m(n)}}M_n u_{m(n)}\Vert$. Thus it is
enough to prove that
\be \label{eq3.40}
\lim _{n\to \infty}\Vert \chi _n u_{m(n)}\Vert =0,\quad \forall n \, .
\ee
To prove (\ref{eq3.40}), let $H'_\rho =\omega ^{-1}H_\rho$ and $\lambda \in \C
\setminus \sigma (H'_0)$ be fixed. Then 
\be \label{eq3.41}
\Vert \chi _n u_m\Vert ^2 \leq c(\Vert \chi _n R'_o(H'_0 -H'_{\rho _m})u_m \Vert ^2
+\Vert \chi _n R'_0 (H'_{\rho _m}-\lambda)u_m\Vert ^2 ) \, .
\ee 
The second term in the right hand side of (\ref{eq3.41}) tends to zero as $n\to
\infty$, because $R'_0 :=(H'_0 -\lambda)^{-1}$ is compact and $(H'_{\rho
_m}-\lambda)u_m \stackrel{w}{\to}0$. The first term in the right hand side of
(\ref{eq3.41}) can be bounded, up to a constant factor, by 
\be \label{eq3.42}
\Vert R'_0 \chi _n (H'_0 -H'_{\rho _m})u_m\Vert ^2 +\Vert [R'_0,\chi
_n](H'_0 -H'_{\rho _m})u_m\Vert ^2 \, . 
\ee 
Now, the first term in (\ref{eq3.42}) can be bounded as follows
\begin{eqnarray} \label{eq3.43}  
\Vert R'_0 \chi _n (H'_0 -H'_{\rho _m})u_m\Vert ^2 & \leq & c\Vert R'_0\Vert ^2 \int _0
^{2\pi}d\varphi \int _0 ^{2n}|(H'_0 -H'_{\rho_m})u_m|^2 dr
                                                       \nonumber\\
& \leq & c\Vert R'_0\Vert ^2 \int _0 ^{2\pi}d\varphi \int _0 ^{2n}(|\xi (r)|^2
|1-e^{-\frac{2}{3}i\varepsilon _m}|+\rho _m |\xi (r)|^3 )^2 |u_m|^2 dr
                                                                \nonumber\\
& \leq & c'(\varepsilon _m n^2 +\rho _m ^2 n^3 )^2 \Vert u_m\Vert ^2
\end{eqnarray}   
where $\sin (\varepsilon _m)=\rho ^2 _m /R$ by hypothesis. In the second
inequality we have used the fact that, for $r\in (0,2n),\, \xi (r)=\xi _0
(r)=r-2i\eta (r),\, \eta (r)=\eta _0 [1-(1+r^3)^{-1/6}]\leq \eta _0$. Since
(\ref{eq3.43}) tends to zero as $n\to \infty$, let us estimate the second term in
(\ref{eq3.42}):
\be \label{eq3.44}
\Vert [R'_0 ,\chi _n](H'_0 -H'_{\rho _m})u_m\Vert \leq \Vert R'_0 [H'_0 ,\chi
_n]R'_0 (H'_0 -H'_{\rho _m})u_m\Vert \leq c'' n^{-1/4} \, .
\ee 
Indeed, the operator $R'_0 [H'_0,\chi _n]=([\chi _n ,(H'_0)^*](R'_0)^*)^*$, when
applied to the bounded sequence $R'_0 (H'_0 -H'_{\rho _m})u_m$ satisfies the
inequality (\ref{eq3.44}) by (\ref{eq3.36}).
%%%%%%%%%%%%%%%%%%%%%%
\\
\\
%%%%%%%%%%%%%%%%%%%%%%
{\it Proof of Theorem \ref{T3.1}.} Since $\lim _{\rho \to 0^+}H_{\rho}u =H_0 u ,\,
\lim _{\rho \to 0^+}H^* _\rho u =H^* _0 u ,\, \forall u\in C_0 ^{\infty}(\R ^+
\times {\mathbf T}^1)$, we can use Corollary \ref{C3.4}, Proposition \ref{P3.6} and
Lemma \ref{L3.7} in order to apply Theorem A.1 of \cite{DW} which provides the
following stability result:
\begin{itemize}
\item[(i)] if $\lambda \notin \sigma (H_0)$ then $(\lambda -H_\rho)^{-1}$ is
uniformly bounded as $\rho \to 0^+$;
\item[(ii)] if $\lambda \in \sigma (H_0)$ then $\lambda$ is stable with respect to
the family $\{H_\rho\}_{\rho >0}$.
\end{itemize} 
Now the proof of the theorem is a consequence of Remark \ref{R3.2}-(IV). 
%%%%%%%%%%%%%%%%%%%%%%
\\
\\
%%%%%%%%%%%%%%%%%%%%%%
{\it Proof of Theorem \ref{T1.2}.} Taking into account the result obtained in
Theorem \ref{T3.1}, the proofs of Theorems 3.13 and 1.3 of \cite{C-2000} can now be
taken over directly without change in order to prove (a) and (b) respectively. 
%%%%%%%%%%%%%%%%%%%%%%
%%%%%%%%%%%%%%%%%%%%%%
\begin{remark} \label{R3.8}
{\rm
\begin{enumerate}
\item[(1)] For $\beta \in C_R ,E(\beta)$ and $\bar E (\bar \beta)$ are the so called
"upper sum" and "lower sum" respectively of the RSPE (see Remark \ref{R.A3} below),
while the distributional Borel sum is given by $f(\beta)=\frac{1}{2}(E(\beta)+\bar E
(\bar \beta))$ and $d(\beta)=E(\beta)-\bar E (\bar \beta)$ is the discontinuity
with zero asymptotic expansion. The result obtained in Theorem \ref{T1.2} can
be interpreted in terms of resonances of the problem as explained in Remark 3.14 of
\cite{C-2000}.
\item[(2)] Similar results can be obtained if we now start from $\Im \beta <0$,
instead of $\Im \beta >0$. We can establish a relationship between the resonance
$E_1 (\beta)$ obtained in this case and $E(\beta)$ following again \cite{C-2000}.
Indead we have $E_1(\beta)=\bar E (\beta)$ for $\beta \in \R$.   
\end{enumerate}
}
\end{remark}
%%%%%%%%%%%%%%%%%%%%%%
\par
\appendix
%%%%%%%%%%%%%%%%%%%%%%%%%%%%%%%%%%%%%
%%%%%%%%%%%%%%%%%%%%%%%%%%%%%%%%%%%%%
\section{Appendix}
\setcounter{equation}{0}%
\setcounter{theorem}{0}% 
%%%%%%%%%%%%%%%%%%%%%%%%%%%%%%%%%%%%%
%%%%%%%%%%%%%%%%%%%%%%%%%%%%%%%%%%%%%
To make the paper self-contained, in this appendix we first recall the notion of
distributional Borel-Leroy summability of order $q$ as introduced in \cite{DBS}.
\begin{definition} \label{D.A1}
Let $q$ be a rational number, $(a_s)_{s\in \N}$ a sequence of real numbers and
$R>0$. We say that the formal series $\sum _{s=0}^{\infty}a_s \beta ^s$ is Borel-
Leroy summable of order $q$ in the distributional sense to $f(\beta)$ for
$0<\beta<R$ if the following conditions are satisfied.
\begin{itemize}
\item[(a)] Set
\be \label{eq.A1}
B(t)=\sum _{s=0}^{\infty}\frac{a_s}{\Gamma (qs +1)}t^s \, .
\ee
Then $B(t)$ is holomorphic in some circle $|t|<\Lambda$; moreover $B(t)$ admits a
holomorphic continuation to the intersection of some neighborhood of
$\R ^+ :=\{t\in \R:\,t>0\}\;\;\mbox{with}\;\; \C^+ :=\{t\in \C:\, \Im t>0\}$.
\item[(b)] The boundary value distribution $B(t+i0)$ exists $\forall t\in \R ^+$,
and the following representation holds:
\be \label{eq.A2}
f(\beta)=\frac{1}{q\beta}\int _0
^{\infty}PP(B(t))e^{-(t/\beta)^{1/q}}\left(\frac{t}{\beta}\right)^{-1+1/q}dt
\ee
for $\beta$ belonging to the Nevanlinna disc of the $\beta ^{1/q}-$plane $C_R
:=\{\beta :\Re \beta ^{-1/q}>R^{-1}\}$, where
$PP(B(t))=\frac{1}{2}(B(t+i0)+\overline{B(t+i0)})$.\\
If $q=1$ the series is called Borel summable in the distributional sense to
$f(\beta)$.  
\end{itemize}
\end{definition}
%%%%%%%%%%%%%%%%%%%%
Let us now recall the criterion for the distributional Borel-Leroy summability
(see \cite{DBS}). As for the ordinary Borel sum, it shows that the representation
(\ref{eq.A2}) is unique among all real functions admitting the prescribed formal
power series expansion and fulfilling suitable analyticity requirements and remainder
estimates. For the sake of simplicity we limit ourselves to the case
$q=1$.
%%%%%%%%%%%%%%%%%%%%
%%%%%%%%%%%%%%%%%%%%
\begin{theorem} \label{T.A2}
Let $f(\beta)$ be bounded and analytic in the Nevanlinna disc $C_R =\{\beta
:\,\Re \beta ^{-1}>R^{-1}\}$ and let $f(\beta)=(\Phi (\beta)-\bar \Phi (\bar
\beta))/2$, with $\Phi (\beta)$ analytic in $C_R$ and such that
\be \label{eq.A3}
\left|\Phi (\beta)-\sum _{s=0}^{N-1}a_s \beta ^s\right|\leq A\sigma (\varepsilon)^N
N!|\beta|^N ,\quad \forall N=1,2,\ldots
\ee
uniformly in $C_{R,\varepsilon}=\{\beta \in C_R :\,\arg \beta \geq -\pi
/2+\varepsilon\},\,\forall \varepsilon >0$. Then the series $\sum
_{s=0}^{\infty}(a_s /s!)u^s$ is convergent for small $|u|$ and it admits an
analytic continuation $B(u)=B_1 (u)+B_2 (u)$, where $B_1 (u)$ is analytic in $C_d ^1
=\{u:\,\mbox{dist}\,(u,\R ^+)<d^{-1}\}$ and $B_2 (u)$ is analytic in $C_d
^{2}=\{u:\, (\Im u>0,\Re u>-d^{-1})\,\mbox{or}\, |u|<d^{-1}\}$ for some $d>0$.
$B(u)$ satisfies
\be \label{eq.A4}
|B(t+i0)|\leq A'(\eta _0)^{-1}e^{t/R}
\ee 
uniformly for $t>0$, for any $\eta _0$ such that $0<\eta _0 <d^{-1}$. Moreover,
setting $PP(B(t))=(B(t+i0)+\overline{B(t+i0)})/2 ,\,f(\beta)$ admits the integral
representation
\be \label{eq.A5}
f(\beta)=\beta ^{-1}\int _0 ^{\infty}PP(B(t))e^{-t/\beta}dt,\qquad \beta \in C_R  
\ee 
i.e. $f(\beta)$ is the distributional Borel sum of $\sum _{s=0}^{\infty}a_s \beta
^s$ for $0<\beta <R$ in the sense of Definition \ref{D.A1}. Conversely, if
$B(u)=\sum _{s=0}^{\infty}(a_s /s!)u^s$ is convergent for $|u|<d^{-1}$ and admits
the decomposition $B(u)=B_1 (u)+B_2 (u)$ with the above quoted properties, then
the function defined by (\ref{eq.A5}) is real-analytic in $C_R$ and $\Phi
(\beta)=\beta ^{-1}\int _0 ^{\infty}B(t+io)e^{-t/\beta}dt$ is analytic and
satisfies (\ref{eq.A3}) in $C_R$. 
\end{theorem}
%%%%%%%%%%%%%%%%%%%%
%%%%%%%%%%%%%%%%%%%%
\begin{remark} \label{R.A3}
{\rm The function $\ds \phi(\beta) = \beta^{-1}\int_0^{\infty}B(t+i0)e^{-t/\beta}dt$
is called "the upper sum" and $\ds \overline{\phi(\overline{\beta})} =
\beta^{-1}\int_0^{\infty}\overline{B(t+i0)}e^{-t/\beta}dt$ "the lower sum" of the
series. It follows that, for $\beta>0$, $f(\beta) = \re{\phi(\beta)}$. On the other
hand with this method we can single out a unique function with zero asymptotic
power series expansion, that is the "discontinuity"
$$   
d(\beta) = \beta^{-1}\int_0^{\infty}(B(t+i0) - \overline{B(t+i0)})e^{-t/\beta}dt =
\phi(\beta) - \overline{\phi(\overline{\beta})}\,.
$$
Thus, $d(\beta) = 2i\im{\phi(\beta)}$, for $\beta>0$.}
\end{remark}
%%%%%%%%%%%%%%%%%%%%

%%%%%%%%%%%%%%%%%%%%%%%%%%%%%%%%%%%%%
%%%%%%%%%%%%%%%%%%%%%%%%%%%%%%%%%%%%%
\section{Appendix}
\setcounter{equation}{0}%
\setcounter{theorem}{0}% 
%%%%%%%%%%%%%%%%%%%%%%%%%%%%%%%%%%%%%
%%%%%%%%%%%%%%%%%%%%%%%%%%%%%%%%%%%%%
                 
{\it Proof of Lemma \ref{L2.1}.} We shall proof the following estimate,
equivalent to (\ref{eq2.1}):
\be \label{eq.B1}
\Vert \Delta u\Vert ^2 +|\sigma|^2 \Vert x^2 u\Vert +|\beta|^2 \Vert Vu\Vert ^2
\leq a\Vert (-e^{-i\alpha}\Delta +\sigma x^2 +|\beta|V)u\Vert ^2 +b\Vert u\Vert ^2
\ee   
$\forall u\in D(H_0)\cap D(V)$, with $\sigma =\gamma e^{-i\alpha}$ varying in
a compact subset of $\{\sigma \in \C \setminus \{0\}:\,|\sigma|^2 > 4|\beta|\sin
\alpha ,\,-\pi <\arg \pi<0\}$. From now on we shall use the notation $-\Delta
=p_1 ^2 +p_2 ^2$, where $p_j =-\frac{\partial ^2}{\partial x_j ^2},\;j=1,2$. As
quadratic forms on $C_0 ^{\infty}(\R^2)\otimes C_0 ^{\infty}(\R^2)$ we have
\begin{eqnarray} \label{eq.B2}
\lefteqn{ (-e^{i\alpha}\Delta +\bar \sigma x^2 +|\beta|V(x))(-e^{-i\alpha}\Delta +\sigma
x^2 +|\beta|V(x)) }&&
                    \nonumber\\
&&=(-e^{i\alpha}\Delta +|\beta|V(x))(-e^{-i\alpha}\Delta +|\beta|V(x))+|\sigma|^2 |x|^4 
                                                                     \nonumber\\
&&\quad+\Re \sigma [(-e^{i\alpha}\Delta +|\beta|V(x))x^2 +x^2
(-e^{-i\alpha}\Delta +|\beta|V(x))]
                                  \nonumber\\
&&\quad+i\Im \sigma [(-e^{i\alpha}\Delta +|\beta|V(x))x^2 -x^2 (-e^{-i\alpha}\Delta
+|\beta|V(x))]
             \nonumber\\
&&=\bigg|\frac{\Re \sigma}{\sigma}\bigg|(-e^{-i\alpha}\Delta +|\beta|V(x)\pm
|\sigma|x^2)(-e^{-i\alpha}\Delta +|\beta|V(x)\pm |\sigma|x^2)
                                                          \nonumber\\
&&\quad+\left(1-\left|\frac{\Re \sigma}{\sigma}\right|\right)[(-e^{i\alpha}\Delta
+|\beta|V(x))(-e^{-i\alpha}\Delta +|\beta|V(x))+|\sigma|^2 |x|^4]
                                                               \nonumber\\
&&\quad+i\Im \sigma (-e^{i\alpha}\Delta x^2 +e^{-i\alpha}x^2 \Delta)
                                                                  \nonumber\\
&&\geq \left(1-\left|\frac{\Re \sigma}{\sigma}\right|\right)[(-e^{i\alpha}\Delta
+|\beta|V(x))(-e^{-i\alpha}\Delta +|\beta|V(x))+|\sigma|^2 |x|^4]
                                                              \nonumber\\
&&\quad+i\Im \sigma \cos \alpha [-\Delta ,x^2]-\Im \sigma \sin \alpha (-\Delta x^2 -x^2
\Delta)
      \nonumber\\
&&=\left(1-\left|\frac{\Re \sigma}{\sigma}\right|\right)[\cdots]+2\Im \sigma \cos \alpha
(p_1 x_1 +x_1 p_1 +p_2 x_2 +x_2 p_2)
                                   \nonumber\\
&&\quad-\Im \sigma \sin \alpha (p_1 ^2 x_1 ^2 +x_1 ^2 p_1 ^2 +p_2 ^2 x_2 ^2 +x_2 ^2 p_2
^2 +2x_1 ^2 p_2 ^2 +2x_2 ^2 p_1 ^2)
                                  \nonumber\\
&&\mbox{(since $\Im \sigma <0,\sin \alpha >0$ and $x_1 ^2 p_2 ^2 +x_2 ^2 p_1 ^2 \geq 0$)}
                                                                                       \nonumber\\
&&\geq \left(1-\left|\frac{\Re \sigma}{\sigma}\right|\right)[\cdots]+2\Im \sigma
\cos \alpha (p_1 x_1 +x_1 p_1 +p_2 x_2 +x_2 p_2)
                                               \nonumber\\
&&\quad-\Im \sigma \sin \alpha (p_1 ^2 x_1 ^2 +x_1 ^2 p_1 ^2 +p_2 ^2 x_2 ^2 +x_2 ^2 p_2
^2)
\nonumber\\
&&=\left(1-\left|\frac{\Re \sigma}{\sigma}\right|\right)[\cdots]\pm 2\Im \sigma |\cos
\alpha|(p_1 x_1 +x_1 p_1 +p_2 x_2 +x_2 p_2)
                                         \nonumber\\
&&\quad-\Im \sigma \sin \alpha (-4+2p_1 x_1 ^2 p_1 +2p_2 x_2 ^2 p_2)
                                                            \nonumber\\
&&\mbox{(since $p_j x_j ^2 p_j \geq 0,j=1,2$)}
                                           \nonumber\\
&&\geq \left(1-\left|\frac{\Re \sigma}{\sigma}\right|\right)[\cdots]-2\Im \sigma |\cos
\alpha|[(p_1 \mp x_1)^2 -p_1 ^2 -x_1 ^2 +(p_2 \mp x_2)^2 -p_2 ^2 -x_2 ^2]
                                                                       \nonumber\\ 
&&\quad+4\Im \sigma \sin \alpha 
                          \nonumber\\
&&\geq \left(1-\left|\frac{\Re \sigma}{\sigma}\right|\right)[\cdots]+2\Im \sigma |\cos
\alpha|(p_1 ^2 +x_1 ^2 +p_2 ^2 +x_2 ^2)+4\Im \sigma \sin \alpha \, .
\end{eqnarray}
We shall prove below that the term inside square brackets in (\ref{eq.B2}) satisfies the
following estimate:
\begin{eqnarray} \label{eq.B3}
\lefteqn{ (-e^{i\alpha}\Delta +|\beta|V(x))(-e^{-i\alpha}\Delta +|\beta|V(x))+|\sigma|^2
|x|^4 }&&\nonumber\\
&&\geq a_1 [(p_1 ^2 +p_2 ^2)^2 +|\beta|^2 V(x)^2]+\frac{|\sigma|^2}{2}|x|^4 -b_1 .
\end{eqnarray}
for suitable constants $a_1 ,b_1 >0$, independent of $\gamma \in \Omega$ and $\alpha$ in
a compact subset of $]0,\pi[$.\\
Now, using (\ref{eq.B3}) and setting $A=(1-\left|\frac{\Re \sigma}{\sigma}\right|)a_1
,B=\frac{1}{2}(1-\left|\frac{\Re \sigma}{\sigma}\right|)$ and $b_2 =(1-\left|\frac{\Re
\sigma}{\sigma}\right|)b_1 -4\Im \sigma \sin \alpha$, (\ref{eq.B2}) can be bounded from
below by:
\begin{eqnarray} \label{eq.B4}
\lefteqn{ A[(p_1 ^2 +p_2 ^2)^2 +|\beta|^2 V(x)^2]+B|\sigma|^2 |x|^4 -b_2 }&&
                                                                          \nonumber\\
&&+2\Im \sigma |\cos \alpha|(p_1 ^2 +x_1 ^2 +p_2 ^2 +x_2 ^2)
                                                          \nonumber\\
&&=\left[Aa'(p_1 ^2 +p_2 ^2)^2 +2\Im \sigma |\cos \alpha|(p_1 ^2 +p_2 ^2)-b_2
+\frac{b'}{2}\right]
                  \nonumber\\
&&\quad+A|\beta|^2 V(x)^2 +\left[a'|B||\sigma|^2 |x|^4 +2\Im \sigma |\cos \alpha|(x_1 ^2
+x_2 ^2)+\frac{b'}{2}\right]
                          \nonumber\\
&&\quad+A(1-a')(p_1 ^2 +p_2 ^2)^2 +B(1-a')|\sigma|^2 |x|^4 -b' \, .
\end{eqnarray}
Since the terms inside square brackets in (\ref{eq.B4}) are positive for a suitable
choice of the constants $a',b'>0, a'<1$, we finally obtain 
\begin{eqnarray} \label{eq.B5}
\lefteqn{ (-e^{i\alpha}\Delta +\bar \sigma x^2 +|\beta|V(x))(-e^{-i\alpha}\Delta +\sigma
x^2 +|\beta|V(x)) }&&
                    \nonumber\\
&&\geq A(1-a')(p_1 ^2 +p_2 ^2)^2 +A|\beta|^2 V(x)^2 +B(1-a')|\sigma|^2 |x|^4 -b' \, .
\end{eqnarray}
Now (\ref{eq.B1}) follows from (\ref{eq.B5}) with $a=\min (A(1-a'),B(1-a'))$ and
$b=\frac{b'}{a}$. In order to complete the proof of the lemma we need to prove
(\ref{eq.B3}). We have 
\begin{eqnarray} \label{eq.B6}
\lefteqn{ (-e^{i\alpha}\Delta +|\beta|V(x))(-e^{-i\alpha}\Delta
+|\beta|V(x))+\frac{|\sigma|^2}{2}|x|^4 }&&
                                                    \nonumber\\
&&=(p_1 ^2 +p_2 ^2)^2 +|\beta|^2 V(x)^2 +|\beta|\cos \alpha [(p_1 ^2 +p_2
^2)V(x)+V(x)(p_1 ^2 +p_2 ^2)] 
                            \nonumber\\
&&\quad+i|\beta|\sin \alpha [(p_1 ^2 +p_2 ^2),V(x)]+\frac{|\sigma|^2}{2}|x|^4
                                                                   \nonumber\\
&&=(p_1 ^2 +p_2 ^2)^2 +|\beta|^2 V(x)^2 \pm |\beta||\cos \alpha|[(p_1 ^2 +p_2
^2)V(x)+V(x)(p_1 ^2 +p_2 ^2)]
                           \nonumber\\
&&\quad-2\sin \alpha |\beta|\left[x_2 (p_1 x_1 +x_1 p_1)+x_1 (p_2 x_2 +x_2 p_2)
+\frac{1}{2}(x_2 ^2 p_2 +p_2 x_2 ^2)\right]+\frac{|\sigma|^2}{2}|x|^4
                                          \nonumber\\
&&=|\cos \alpha|(p_1 ^2 +p_2 ^2 \pm |\beta|V(x))^2 +(1-|\cos \alpha|)[(p_1 ^2 +p_2
^2)^2 +|\beta|^2 V(x)^2]
                      \nonumber\\
&&\quad-2\sin \alpha |\beta|\left[x_2 (p_1 x_1 +x_1 p_1)+x_1 (p_2 x_2 +x_2 p_2)
+\frac{1}{2}(x_2 ^2 p_2 +p_2 x_2 ^2)\right]+\frac{|\sigma|^2}{2}|x|^4
                                                                    \nonumber\\
&&\geq (1-|\cos \alpha|)[(p_1 ^2 +p_2 ^2)^2 +|\beta|^2 V(x)^2]+\frac{|\sigma|^2}{2}|x|^4
+2|\beta|\sin \alpha \bigg[(p_1 -x_1 x_2)^2
                                          \nonumber\\
&&\quad+(p_2 -x_1 x_2)^2+\frac{1}{2}(p_2 -x_2 ^2)^2 -\left(p_1 ^2 +\frac{3}{2}p_2 ^2
-2x_1 ^2 x_2 ^2 -\frac{1}{2}x_2 ^4\right)\bigg]
                                             \nonumber\\
&&\geq (1-|\cos \alpha|)[(p_1 ^2 +p_2 ^2)^2 +|\beta|^2 V(x)^2]-2|\beta|\sin\alpha
\left(p_1 ^2 +\frac{3}{2}p_2 ^2 +2x_1 ^2 x_2 ^2 +\frac{1}{2}x_2 ^4\right)
                                                                        \nonumber\\
&&\quad+\frac{|\sigma|^2}{2}|x|^4
                                \nonumber\\
&&=\left[(1-|\cos \alpha|)a_2 (p_1 ^2 +p_2 ^2)^2 -2|\beta|\sin \alpha \left(p_1 ^2
+\frac{3}{2}p_2 ^2\right)+b_1\right]
                                   \nonumber\\
&&\quad+(1-|\cos \alpha|)(1-a_2)(p_1 ^2 +p_2 ^2)^2 +(1-|\cos \alpha|)|\beta|^2 V(x)^2
-b_1
  \nonumber\\
&&\quad+\frac{|\sigma|^2}{2}|x|^4 -|\beta|\sin \alpha (4x_1 ^2 x_2 ^2 +x_2 ^4).
\end{eqnarray}
Now, for a suitable choice of the constants $0<a_2<1,b_3 >0$, the term in square
brackets in (\ref{eq.B6}) is positive and therefore (\ref{eq.B6}) can be bounded from
below by:
\begin{eqnarray} \label{eq.B7}
\lefteqn{ (1-|\cos \alpha|)(1-a_2)(p_1 ^2 +p_2 ^2) }&&
                                                \nonumber\\
&&+(1-|\cos \alpha|)|\beta|^2 V(x)^2 -b_1 +\frac{|\sigma|^2}{2}|x|^4 -|\beta|\sin \alpha
(4x_1 ^2 x_2 ^2 +x_2 ^4).
\end{eqnarray} 
Next notice that, for $|\beta|\sin \alpha <\frac{|\sigma|^2}{4}$, we have 
\begin{eqnarray*}
\lefteqn{ \frac{|\sigma|^2}{2}|x|^4 -4|\beta|\sin \alpha \, x_1 ^2 x_2 ^2 -|\beta|\sin
\alpha \, x_2 ^4 }&& \\
&&=\frac{|\sigma|^2}{2}x_1 ^4 +\left(\frac{|\sigma|^2}{2}-|\beta|\sin \alpha \right)x_2
^4 +(|\sigma|^2 -4|\beta|\sin \alpha)x_1 ^2 x_2 ^2 \geq 0 \, .
\end{eqnarray*}
Thus, we finally obtain
\begin{eqnarray} \label{eq.B8}
\lefteqn{ (-e^{i\alpha}\Delta +|\beta|V(x))(-e^{-i\alpha}\Delta +|\beta|V(x))+|\sigma|^2
|x|^4 }&&
        \nonumber\\
&&\geq (1-|\cos \alpha|)(1-a_2)(p_1 ^2 +p_2 ^2)^2 +(1-|\cos \alpha|)V(x)^2
                                                                      \nonumber\\
&&\qquad+\frac{|\sigma|^2}{2}|x|^4 -b_1
\end{eqnarray}
which corresponds to (\ref{eq.B3}) with $a_1 =(1-|\cos \alpha|)(1-a_2)$.

\end{document}